\documentclass[aps,prl,twocolumn,showpacs,groupedaddress,amsmath,amssymb]{revtex4}

\usepackage{graphicx}
\usepackage{color}
\usepackage{comment}
\usepackage{bm}
\usepackage{latexsym}

\usepackage[symbol]{footmisc}
\usepackage{hyperref}

\begin{document}

\title{Floquet Protocols of Adiabatic State-Flips and Re-Allocation of Exceptional Points}

\author{Dashiell Halpern, Huanan Li\footnotemark[1], Tsampikos Kottos\footnotemark[1]}

\address{Physics Department, Wesleyan University, Middletown CT-06459, USA}

\date{\today}

\begin{abstract}
We introduce the notion of adiabatic state-flip of a Floquet Hamiltonian associated with a non-Hermitian system that it is subjected to two driving 
schemes with clear separation of time scales. The fast (Floquet) modulation scheme is utilized to re-allocate the exceptional points in the parameter 
space of the system and re-define the topological features of an adiabatic cyclic modulation associated with the slow driving scheme. Such 
topological re-organization can be used in order to control the adiabatic transport between two eigenmodes of the Floquet Hamiltonian. The 
proposed scheme provides a degree of reconfigurability of adiabatic state transfer which can find applications in system control in photonics and 
microwave domains.
\end{abstract}


\maketitle

\footnotetext[1] {Corresponding Authors:hli01@wesleyan.edu;tkottos@wesleyan.edu}

{\it Introduction--} The adiabatic theorem of Hermitian quantum mechanics is at the heart of many phenomena with far reaching technological
applications. In simple terms it states that when a system described by a (sufficiently) slowly varying Hamiltonian $H(t)$ is initially prepared 
at a non-degenerate normal mode of $H(t=t_0)$, it will remain in the corresponding normal mode of the instantaneous $H(t)$ throughout the
evolution. Consequently a cyclic adiabatic change in a multi-parameter space will return the system to its initial state, with possible an overall
phase modification -- the famous Berry phase. The latter turns out to be insensitive to the specifics of the adiabatic motion and depends only 
on the choice of the path in the parameter space \cite{1}. 

The situation is richer when a non-Hermitian system is driven adiabatically. In this case, the existence of non-Hermitian spectral singularities
known as exceptional points (EP) (simultaneous coalesce of eigenvalues and eigenvectors) \cite{2} can lead to a completely different physics 
than the one predicted by the adiabatic theorem of a Hermitian system. If the parametric variation of the Hamiltonian around an EP occurs 
quasi-statically, the instantaneous eigenstates transform into each other at the end of the cycle with only one of them acquiring a geometric 
phase \cite{3,34,4}. If, however, the adiabatic evolution around the EP is dynamical (but still slow) then only one state dominates the output and
what determines this preferred eigenstate is the sense of rotation in the parameter space. This surprising effect has been recently confirmed
in microwave and optomechanical systems. The growing attention to this chiral mode switching (state-flip) has roots in its potential technological 
implications; specifically the robustness of the associated adiabatic transfer against small fluctuations in the control path \cite{8} is an asset 
for many practical applications. 

Less effort has been devoted in proposing driving protocols that alter the dynamics by changing the topological features of a {\it fixed} (both in 
position and direction) adiabatic control path by re-allocating the EP in the parameter space. Here, we explore this viewpoint and provide {\it 
reconfigurable} protocols that manipulate the relative position of an EP singularity, by placing it inside or outside a fixed closed adiabatic control 
path. Then, we harvest such topological re-organization in order to control adiabatic state transfer between two states of a non-Hermitian system. 
The scheme involves a Floquet driving with a (fast) frequency which is rationally related to the inverse period needed for a cyclic adiabatic 
variation associated with two other control parameters. First we show that the Floquet driving re-allocates or even creates/annihilates EPs in the 
parameter space in a controllable manner. Then we introduce the {\it Floquet scenario of adiabatic state-flip}. We show that the ``Floquet" EPs have 
the same topological features as the ones associated with a static non-Hermitian Hamiltonian. Finally we control an adiabatic state-flip from 
one Floquet eigenmode to another by re-allocating (inside or outside the cyclic adiabatic path) the Floquet EP via management of the Floquet 
driving.

{\it Theoretical Modeling --} We consider the following analytically solvable time-dependent non-Hermitian system:
\begin{align}
\imath\frac{d}{dt}|\Psi(t)\rangle= & H(t)|\Psi(t)\rangle;\;H(t)=\begin{bmatrix}a & e^{\imath\omega t}\\
e^{-\imath\omega t}b_{\Omega}(t) & -a
\end{bmatrix}\label{eq: main}
\end{align}
where $b_{\Omega}(t)=e^{\imath\Omega t}$ is a complex time-varying parameter. For simplicity we assume that the constant $a$ is a real 
number. Furthermore, we impose separation of time scales between the two driving schemes by requesting that $\omega=N\Omega$ where 
$N\gg1$ is a positive integer. The adiabatic variation of $b_{\Omega}\left(t\right)$ is ensured by the condition $\Omega\rightarrow0$ while 
the {\it Floquet} (fast) driving of $H(t)$ is controlled by $\omega$ and shall be used to manipulate the position of the EP in the parameter 
space. During the time interval $t\in [0,T_{\Omega}=\frac{2\pi}{\Omega}]$ the parameter $b_{\Omega}$ defines a unit circle centered around 
the origin of the ${\cal R}e(b)-{\cal I}m(b)$ parameter space. A clockwise (CW) motion along the parametric circle occurs whenever $\Omega>0$. 
The case of counter-clockwise (CCW) evolution corresponds to $\Omega<0$. At the end of the circle we have that $H\left(t+T_{\Omega}\right)
=H\left(t\right)$.

{\it Preparation, observation and state-flipping in Floquet Basis--} In the previous adiabatic cyclic schemes the emphasis of the analysis was given 
to the notion of instantaneous Hamiltonians and their corresponding eigenvalues and eigenvectors \cite{5,7}. In contrast, in the presence of Floquet 
(fast) driving, like in Eq. (\ref{eq: main}), the appropriate description of the dynamics is done using the notion of the Floquet Hamiltonian. In this 
case we shall assume that the preparation and the representation of our evolve state(s) occurs in the corresponding Floquet basis. Therefore, any 
notion of state-flip via adiabatic encircle of EPs has to be analyzed and examined with respect to this basis. At the same time any notion of EPs has
to involve information about the Floquet quasi-energies.

\begin{figure}
\includegraphics[width=1\columnwidth,keepaspectratio,clip]{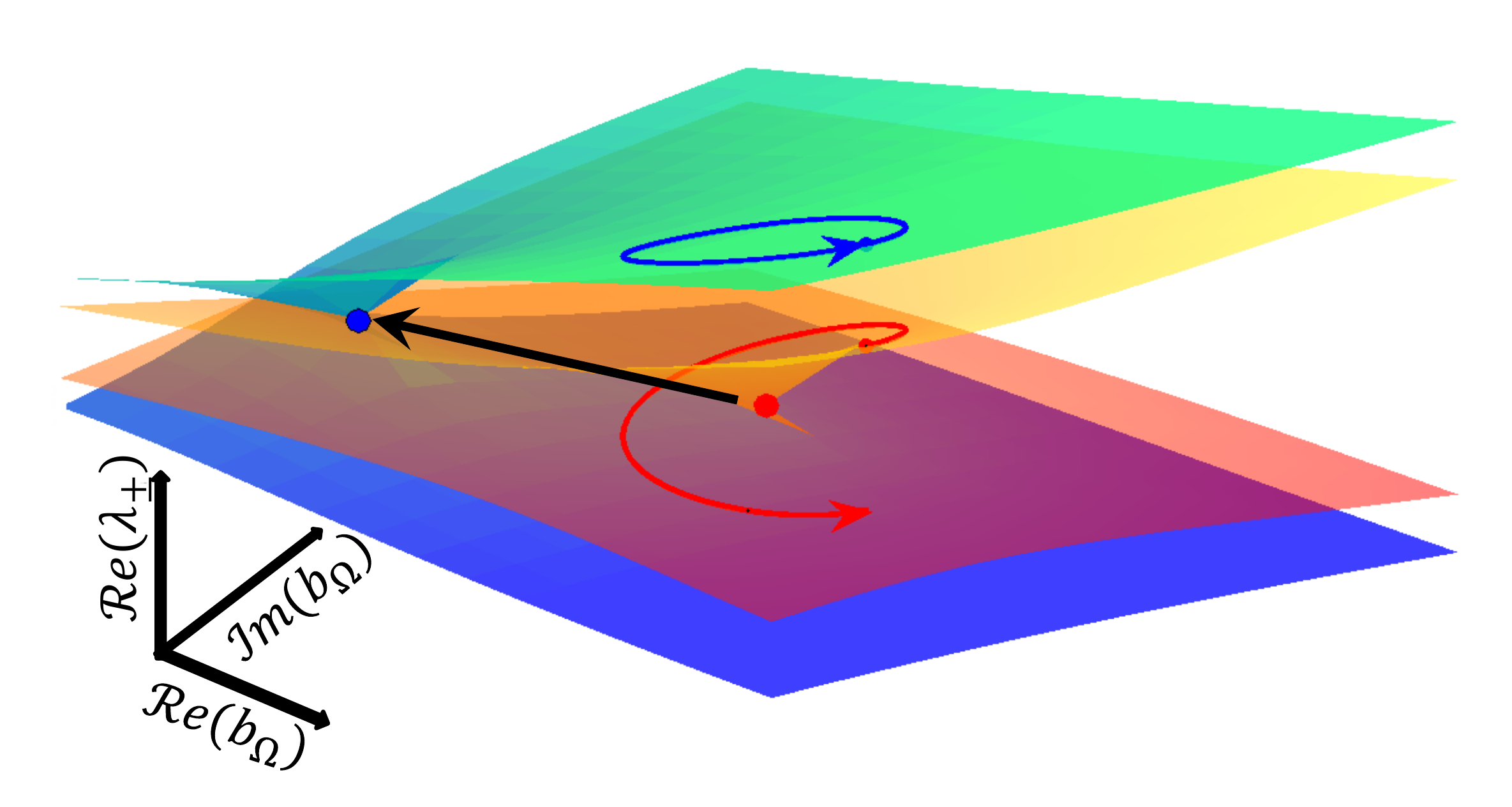}
\begin{center}
 \caption{Overview of the instantaneous Floquet eigenvalue surfaces $\lambda_{\pm}$  versus the adiabatic complex parameter $b_{\Omega}$ 
for the system of Eq.~(\ref{eq: main}). The eigenvalue surfaces correspond to the real part of $\lambda_{\pm}$ for two different Floquet driving 
frequencies $\omega=0$ (inner red and orange surfaces) and $\omega=6$ (outer blue and gree surfaces). The solid red line represents the real
part of $p(t)$, see Eq. (\ref{pt}). The direction of the {\it fixed} adiabatic cyclic variation of $b_{\Omega}$ is chosen in a way that we have state-flip 
for $\omega=0$. When $\omega=6$ the EP is re-allocated (see black arrow) outside the adiabatic cycle and the system remains at the same 
instantaneous Floquet state as the initial preparation. In both case $\left|\Psi(t=0)\right\rangle=\left|\lambda_{+}(t=0)\right\rangle$ and the parameter $a=0$. In addition, we shift $\lambda_{\pm}$ in Eq.~(\ref{eq: eigen}) by a real constant so that $\lambda_{\pm}=0$ at the EPs.  
}
\label{fig1}
\end{center}
\end{figure}

To be specific, we consider a time period of $H\left(t\right)$ (see Eq. (\ref{eq: main})) from $t=0$ to $T_{\Omega}=\frac{2\pi}{\Omega}$. We define
the observation times $t$ to be integer multiples of $2\pi/\omega$. We further assume that, between two consequent observation times, the slow 
varying parameter $b_{\Omega}$ is approximately a constant. Then the corresponding evolution operator $\tilde{U}\left(t+\frac{2\pi}{\omega},t\right)$ 
is \cite{hanggi}
\begin{align}
\tilde{U}\left(t+\frac{2\pi}{\omega},t\right)\equiv e^{-\imath\frac{2\pi}{\omega}H_{F}};\quad
H_{F}= & \begin{bmatrix}a+\omega & 1\\
b_{\Omega} & -a
\end{bmatrix}\label{eq: Floquet H}
\end{align}
where $H_F$ is the ``instantaneous'' Floquet Hamiltonian. Its quasi-energies and eigenvectors are
\begin{align}
\lambda_{\pm}= & \frac{\omega}{2}\pm\sqrt{b_{\Omega}+n^{2}};\quad \left|\lambda_{\pm}\right\rangle = & \left[\begin{array}{c}
n\pm\sqrt{b_{\Omega}+n^{2}}\\
b_{\Omega}
\end{array}\right]\label{eq: eigen}
\end{align}
where $n\equiv a+\frac{\omega}{2}$. When $b_{\Omega}$ is considered time-independent, the eigensystem Eq. (\ref{eq: eigen}) has an EP 
degeneracy at $b^{EP}=-n^2=
-(a+\omega/2)^2$. Depending on the value of the Floquet frequency $\omega$, the {\it Floquet EP} can be inside ($|n|<1$), close ($|n|
\gtrapprox 1$) or far ($|n|\gg 1$) away from the adiabatic unit circle that is defined by the variation of the complex parameter $b_{\Omega}(t)$. 

We define our adiabatic state-flip control protocols in the following way. Assume that at $t=0$ we prepare the system at a Floquet eigenstate 
$\left|\Psi(t=0)\right\rangle=\left|\lambda_{\pm} \right\rangle$. Then we evolve this initial state under Eq. (\ref{eq: main}) where $b_{\Omega}$ 
is a slow-varying complex parameter that performs a {\it fixed} close path in the complex $b_{\Omega}$ parameter space. We want to demonstrate 
that a controlled variation of $\omega$ can lead to evolved states $\left|\Psi(t)\right\rangle$ at $t=T_{\Omega}$ which are proportional to the 
complementary (or the same) Floquet eigenstates $\left|\lambda_{\mp}\right\rangle$ ($\left|\lambda_{\pm}\right\rangle$).

The dynamics, in the limit of distinct time scales $N\gg 1$, can be evaluated numerically by applying the Floquet evolution matrix Eq. (\ref{eq: Floquet H}) 
to the initial preparation $\left|\Psi(t=T_{\Omega})\right\rangle=\prod_{l=0}^{N-1} {\tilde U}\left(2\pi (l+1)/\omega,2\pi l/\omega\right)\left|\Psi(0)\right\rangle$ \cite{Tsironis}.
Note that during the numerical evaluation one needs to consider the variations in $b_{\Omega}$ between subsequent time steps. For the specific
Hamiltonian Eq. (\ref{eq: Floquet H}) the evolved state $\left|\Psi(t)\right\rangle$ can be analytically evaluated. This analysis is described below.


{\it Dynamics--} We want to evaluate the evolution of the state $|\Psi(t)\rangle$ at integer multiples of $2\pi/\omega$ (within the adiabatic cycle $t \in [0, 
T_{\Omega}]$), and specifically its form at the end of the adiabatic circle. First, using a time-depended transformation $U(t)$, we eliminate 
the Floquet driving from Eq.~(\ref{eq: main}). The transformation $U(t)$ is:
\begin{align}
U\left(t\right)= & \left[\begin{array}{cc}
e^{\imath\frac{\omega t}{2}}\left(1+\frac{n^{2}}{b_{\Omega}}\right) & e^{\imath\frac{\omega t}{2}}\frac{n}{b_{\Omega}}\\
0 & e^{-\imath\frac{\omega t}{2}}
\end{array}\right]
\label{eq: transformation}
\end{align}
Substituting $|\Psi(t)\rangle=U\left(t\right)
|\varphi(t)\rangle$ in Eq. (\ref{eq: main}) we get \cite{note1}
\begin{align}
\imath\frac{d}{dt}|\varphi(t)\rangle= & H_{eff}|\varphi(t)\rangle,\:H_{eff}=\left[\begin{array}{cc}
0 & 1\\
b_{\Omega}+n^{2} & 0
\end{array}\right]\label{eq: main2}
\end{align}
where $H_{eff}$ involves the Floquet (fast) frequency $\omega$ as a parameter. Below, without loss of generality, we assume that $n>0$. 
For completeness, we also find the instantaneous eigenvalues and eigenvectors of $H_{eff}$:
\begin{align}
E_{eff}^{\pm}\left(t\right)= & \pm\sqrt{b_{\Omega}+n^{2}};\:\left|E_{eff}^{\pm}\left(t\right)\right\rangle =\left[\begin{array}{c}
1\\
 E_{eff}^{\pm}\left(t\right)
\end{array}\right].\label{emods_eff}
\end{align}
These eigenvectors are related with the Floquet eigenvectors Eq.~(\ref{eq: eigen}) via the time-dependent transformation Eq.~(\ref{eq: transformation}), {\it i.e.,} 
$U(t)\left|E_{eff}^{\pm}(t)\right\rangle \propto\left|\lambda_{\pm}(t)\right\rangle$, when $t$ is an integer multiple of $2\pi/\omega$ (observation times). Similarly the instantaneous modes Eq.~(\ref{emods_eff}) have the same EP singularity as the instantaneous Floquet eigenvalues.

Let us first discuss the CW variation of $b_{\Omega}$. The general solution of Eq.~(\ref{eq: main2}) can be easily expressed in terms of 
modified Bessel functions as \cite{maths} :
\begin{align}
|\varphi(t)\rangle= & C_{1}\left[\begin{array}{c}
I_{\nu}\left(\nu z\right)\\
-e^{\tau}I'_{\nu}\left(\nu z\right)
\end{array}\right]+C_{2}\left[\begin{array}{c}
K_{\nu}\left(\nu z\right)\\
-e^{\tau}K'_{\nu}\left(\nu z\right)
\end{array}\right]\label{eq: Gsolution}
\end{align}
where $C_{1}$ and $C_{2}$ are arbitrary constants determined by the initial conditions, $I_{\nu}\left(\nu z\right)$ and $K_{\nu}\left(\nu z\right)$ 
are the $\nu(=\frac{2n}{\Omega})$-th order modified Bessel function of the first and second kind, $I'_{\nu}\left(\nu z\right)$ ($K'_{\nu}\left(\nu z\right)$) 
is the derivative of $I_{\nu}\left(\nu z\right)$ ($K_{\nu}\left(\nu z\right)$) with respect to the argument and $z=\sqrt{b_{\Omega}}/n$, $\tau=\imath\Omega t/2$. Note that the 
general solution in Eq.~(\ref{eq: Gsolution}) is not a periodic function since the modified Bessel function $I_{\nu}$ 
and $K_{\nu}$ are multivalued functions. 

We are mainly interested in the form of Eqs.~(\ref{eq: Gsolution}) at the beginning $t=0$ and at the end
$t=T_{\Omega}$ of the adiabatic circle. For $t=0$, and under the adiabaticity condition $\Omega\rightarrow 0^+$ (corresponding to $\nu={2n\over 
\Omega}\rightarrow+\infty$), we can easily show that appropriate choice of $C_1,C_2$ leads to the forms
\begin{equation}
|\varphi(t=0)\rangle \propto
\left\{
\begin{array}{c}
\left|E_{eff}^{-}(0)\right\rangle  {\rm when} \,\,C_{1}=1; C_{2}=0 \\
\left|E_{eff}^{+}(0)\right\rangle {\rm when} \,\,C_{1}=0; C_{2}=1
\end{array}
\right.
\label{initial}
\end{equation}
where we have taken into account that $I'_{\nu}\left(\frac{\nu}{n}\right)/I{}_{\nu}\left(\frac{\nu}{n}\right)\sim\sqrt{1+n^{2}}$ and that $K'_{\nu}\left
(\frac{\nu}{n}\right)/K{}_{\nu}\left(\frac{\nu}{n}\right)\sim-\sqrt{1+n^{2}}$ \cite{maths}. 

Next we evaluate the evolved state Eq.~(\ref{eq: Gsolution}) at the end of the adiabatic circle $t=T_{\Omega}$. Using the identities $I_{\nu}
\left(ze^{\imath m\pi}\right)= e^{\imath\nu m\pi}I_{\nu}\left(z\right)$ and $K_{\nu}\left(ze^{\imath m\pi}\right)= e^{-\imath\nu m\pi} K_{\nu}\left(z\right)
-\imath\pi\sin\left(\nu m\pi\right)\csc\left(\nu\pi\right)I_{\nu}\left(z\right)$ ($m$ is an arbitrary integer) \cite{maths}, Eq.~(\ref{eq: Gsolution}) can be
written, at $t=T_{\Omega}$, as
\begin{align}
 |\varphi(T_{\Omega})\rangle
= &\left(C_{1}e^{\imath\nu\pi}-\imath\pi C_{2}\right)\left[\begin{array}{c}
I_{\nu}\left(\frac{\nu}{n}\right)\\
-I'_{\nu}\left(\frac{\nu}{n}\right)
\end{array}\right]\nonumber \\
&+C_{2}e^{-\imath\nu\pi}\left[\begin{array}{c}
K_{\nu}\left(\frac{\nu}{n}\right)\\
-K'_{\nu}\left(\frac{\nu}{n}\right)
\end{array}\right].\label{eq: endingP}
\end{align}

In the adiabatic limit $\Omega\rightarrow 0^{+}$, we can approximate both 
$I_{\nu}\left(\frac{\nu}{n}\right)$ and $I'_{\nu}\left(\frac{\nu}{n}\right)$ using their asymptotic forms which are dominated by the exponential 
factor $e^{\nu\eta}$ \cite{maths}. Similarly $K\left(\frac{\nu}{n}\right)$ and $K'_{\nu}\left(\frac{\nu}{n}\right)$ are dominated by the factor 
$e^{-\nu\eta}$ \cite{maths}. In all cases $\eta\equiv\sqrt{1+1/n^{2}}-\ln\left(n+\sqrt{1+n^{2}}\right)$. Most importantly, its sign is controlled 
by the magnitude of the Floquet driving frequency $\omega$ via the parameter $n$. At this point, it is important to remind that $n$ also 
defines the position of the Floquet EPs (see Eq.~(\ref{emods_eff}) and discussion below). We find that the transition from positive definite 
$\eta$ to negative $\eta$-values occurs at $n_C\approx 1.51$.

For $\eta>0$, corresponding to $n< n_C$, we get that irrespective of the initial conditions Eq. (\ref{initial}) the final state $|\varphi(T_{\Omega})
\rangle$ is dominated by the first term in Eq.~(\ref{eq: endingP}) \cite{note2}. Taking into account that $I'_{\nu}\left(\frac{\nu}{n}\right)/I{}_{\nu}
\left(\frac{\nu}{n}\right)\sim\sqrt{1+n^{2}}$ \cite{maths}, we eventually have that $|\varphi(T_{\Omega})\rangle\propto\left|E_{eff}^{-}\left(0\right)
\right\rangle$.

When $\eta<0$, corresponding to  $n>n_{C}$, one needs to distinguish between two cases for the general solution Eqs.~(\ref{eq: Gsolution},
\ref{eq: endingP}). When $C_{1}=1$ and $C_{2}=0$ (corresponding to $|\varphi(t=0)\rangle\propto \left|E_{eff}^{-}\left(0\right)\right\rangle$, see
Eq. (\ref{initial})), we have that $|\varphi(t=T_{\Omega})\rangle\propto \left|E_{eff}^{-}\left(0\right)\right\rangle$. If on the other hand 
$C_{1}=0$ and $C_{2}=1$ (corresponding to $|\varphi(t=0)\rangle\propto \left|E_{eff}^{+}\left(0\right)\right\rangle$, see Eq. (\ref{initial})),
then we have that $|\varphi(T_{\Omega}) \rangle\propto \left|E_{eff}^{+}\left(0\right)\right\rangle $. Comparison with Eqs. (\ref{initial}) lead us to 
the conclusion that, whenever $n>n_{C}$, we always come back at the initial instantaneous state at the end of the adiabatic circle. 

The case of counter-clockwise (CCW) variation of $b_{\Omega}$ corresponds to the limit of $\Omega\rightarrow0^{-}$ and can be treated 
in a similar manner. Redefining $\nu$ to be $-2n/\Omega$ and performing the same analysis as above we arrive at the following conclusions: 
when $n<n_{C}$, irrespective of the initial preparation Eq. (\ref{initial}), we get  $|\varphi(T_{\Omega})\rangle\propto\left|E_{eff}^{+}\left(0\right)
\right\rangle $; when $n>n_{C}$ we always come back, at the end of the circle $t=T_{\Omega}$, to the initial Floquet eigenstate. 

{\it Floquet state-flip protocols--}The chiral state-flip has been recently predicted theoretically \cite{8, 5} and demonstrated experimentally 
\cite{6, doppler} for adiabatic cyclic variations of non-Hermitian Hamiltonians which encircle an EP-- though these investigation have been performed 
in the absence of any Floquet driving i.e. $\omega=0$. In fact, these studies underplay (or even disregarded) the fact that a state-flip can also 
occur in the case that the EP is outside, but still in the vicinity, of an adiabatic cyclic variation -- as indicated by our analysis above \cite{note3}. 
It is therefore tempting to speculate that the constraint $n^2\equiv |b_{\Omega}^{EP}|\le |b_{\Omega}|=1$ is very restrictive (a sufficient condition) 
i.e. under this condition one {\it necessarily} has state-flip. In fact a less restrictive condition is that $|n|<|n_C|$, (where in our case $1<|n_C|$). 

Let us finally demonstrate that the ability to engineer the topological features of an adiabatic circle via Floquet frequency $\omega$-variations 
can be utilized for the controlled manipulation of state-flip between the two ``instantaneous" Floquet eigenstates $\left|\lambda_{\pm}(t=0)\right
\rangle$. For our analysis we have introduced the measure $p(t)$ which quantifies the relative weight with which each instantaneous eigenvalue
/eigenvector participate in the evolution. Specifically $p(t)$ is defined as
\begin{align}
p(t)={|a_+(t)|^2\lambda_{+}(t)+|a_-(t)|^2\lambda_{-}(t)\over |a_+(t)|^2+|a_-(t)|^2}
\label{pt}
\end{align}
where the eigenvector populations $a_{\pm}(t)$ are evaluated via the decomposition of the evolved state $|\Psi(t)\rangle=U(t)|\varphi(t)\rangle$ in 
the instantaneous Floquet basis Eq. (\ref{eq: eigen}).

In Fig.~\ref{fig1} we report the real part of the instantaneous Floquet eigenmodes $\lambda_{+}$ (upper two surfaces) and $\lambda_{-}$ 
(lower two surfaces) in the complex $b_{\Omega}$-parameter space for two different Floquet driving frequencies $\omega$ when $a=0$. Specifically, the 
red-orange (inner) surfaces correspond to $\omega=0$ while the blue-green (outer) surfaces correspond to $\omega=6$. The projection of the 
real part of $p(t)$ 
in this space is also shown with red (blue) lines for $\omega=0$ ($\omega=6$). Finally the corresponding EP are indicated with filled red (blue) 
circles respectively. We see that the Floquet driving has re-allocated the position of the EP in the $b_{\Omega}$ parameter space. Specifically,
while for $\omega=0$ the EP is inside the adiabatic cyclic path, it is re-allocated far away from the circle when the Floquet frequency is $\omega=6$.
In both cases the variation of $b_{\Omega}$ (both variation rate and direction of variation of control parameters ${\cal R}e(b_{\Omega})-{\cal I}m
(b_{\Omega})$) has been kept fixed. The direction of the adiabatic circle has been chosen in such a way that the system undergoes a state-flip at 
$\omega=0$. When the Floquet frequency has been reconfigured to the value $\omega=6$ the EP has been re-allocated outside the adiabatic 
circle -- thus enforcing the system to evolve to the initial state at the end of the adiabatic cycle $t=T_{\Omega}$.

{\it Other examples--}The above scheme is not specific to the toy model Eq. (\ref{eq: main}). To further confirm its validity, we now perform 
simulations with a driven Hamiltonian which describes two (evanescently) coupled resonators. 
The Hamiltonian takes the form 
\begin{align}
\label{system}
H\left(t\right)= & \left[\begin{array}{cc}
\epsilon_1-\frac{\imath\gamma}{2} & \kappa\\
\kappa & \epsilon_2+\frac{\imath\gamma}{2}
\end{array}\right],
\end{align}
where $\kappa$ is the coupling strength between the two resonators, $\epsilon_1, \epsilon_2$ are the eigenfrequencies of each resonator, and 
$\gamma$ is the gain (loss) parameter that describes the loss (gain) at first (second) resonator. We further assume that the resonant frequencies 
of each of these resonators are periodically modulated as $\epsilon_1(t)=-\epsilon_2(t)= -\frac{F}{2}\sin\omega t-r$. The ``fast" (Fourier) variation 
with period $\omega$ can be achieved via modulation of the permittivities (say via a current injection) of the resonators. The parameters $r$ and 
$\kappa$ represent two additional variables that vary slowly in time (adiabatic parameters). A possible way to achieve this slow modulation is by 
bringing in the vicinity of the resonators a mechanical cantilever which oscillates with a slow frequency $\Omega\ll \omega$.  We note that the same model  Eq.~\ref{system} can be also realized  in the framework of optical coupler \cite{exa}.

Following the same analysis as previously, we first identify the instantaneous Floquet eigenstates and quasi-energies, that will be used for the 
preparation and observation of the evolved state. The associated  ``instantaneous'' Floquet Hamiltonian $H_F$ at times which are multiples of 
$2\pi/\omega$, and in the $\omega\rightarrow \infty$ limit and $F\sim\mathcal{O}\left(\omega\right)$, is \cite{highw}
\begin{align}
H_{F}\approx & \left[\begin{array}{cc}
-r_{\Omega}-\frac{\imath\gamma}{2} & \kappa_{\Omega}J_{0}\left(\frac{F}{\omega}\right)e^{-\imath\frac{F}{\omega}}\\
\kappa_{\Omega}J_{0}\left(\frac{F}{\omega}\right)e^{\imath\frac{F}{\omega}} & r_{\Omega}+\frac{\imath\gamma}{2}
\end{array}\right],
\label{floquet_num}
\end{align}
where $J_{0}$ is the $0$-order bessel function of the first kind. The instantaneous Floquet eigenvalues are evaluated as $\lambda_{\pm}
=\pm\sqrt{\left[\kappa_{\Omega} J_{0}\left(\frac{F}{\omega}\right)\right]^{2}+\left[r_{\Omega}+\imath\gamma/2\right]^{2}}$. The instantaneous 
EPs occurs at $\left(r_{\Omega}^{EP},\kappa_{\Omega}^{EP}\right)=\left(0,\pm\frac{\gamma}{2J_{0}\left(F/\omega\right)}\right)$. The corresponding 
eigenvectors are denoted as $|\lambda_{\pm}(t)\rangle$. Their expressions are rather complicated and we do not give them here. Below,
we have evaluated them numerically for each observation time $t$ during the evolution, via a direct diagonalization of the Hamiltonian Eq. (\ref{floquet_num}).

\begin{figure}
\includegraphics[width=1\columnwidth,keepaspectratio,clip]{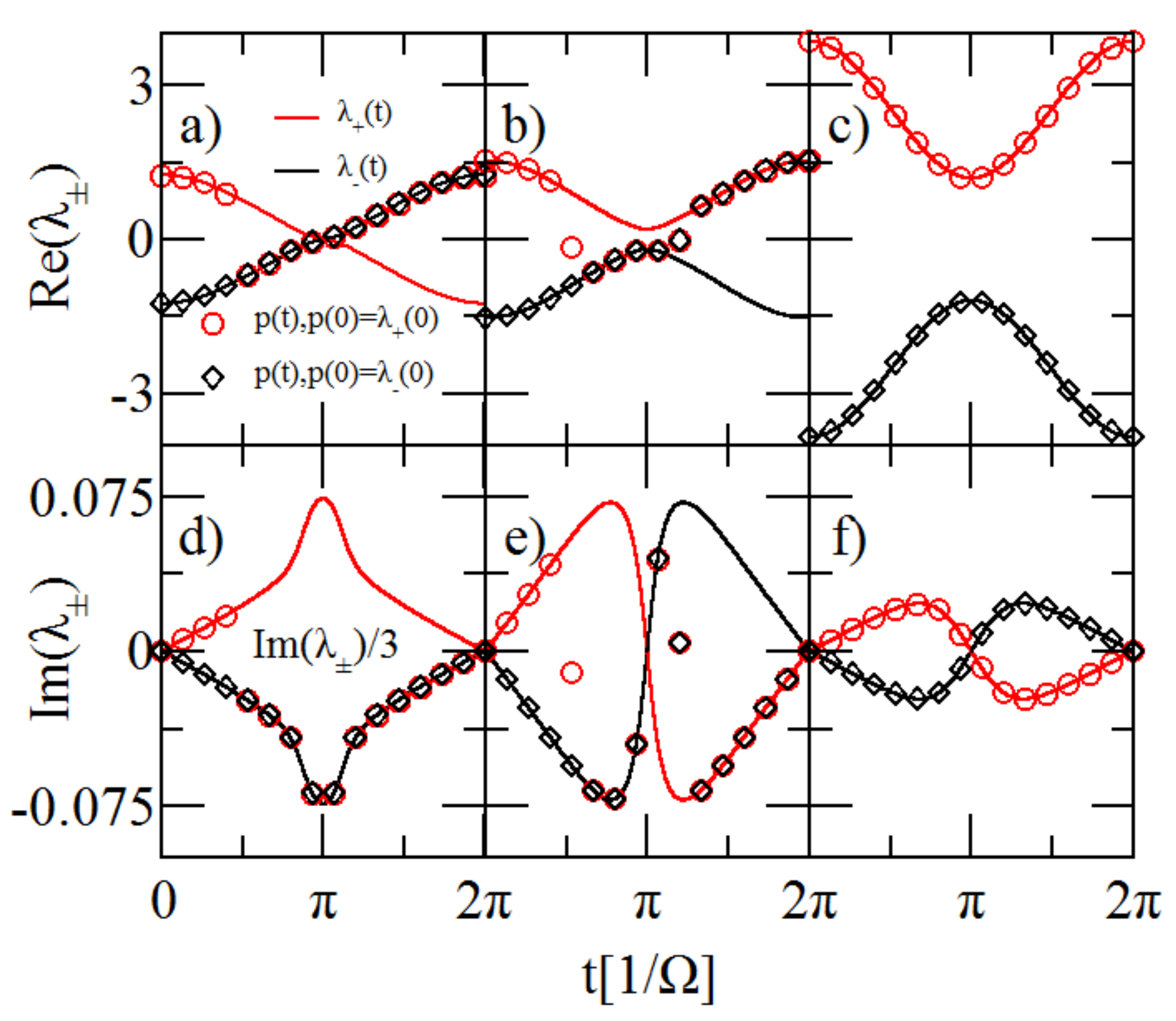}
\begin{center}
\caption{The evolution, during one adiabatic cycle, of the instantaneous Floquet eigenvalues and the associated $p(t)$ for the system described 
by Eq. (\ref{system}). Upper row reports the real part of these quantities while the lower row reports their corresponding imaginary part. (a,d) The 
Floquet frequency is $\omega=50$ and it is chosen in a way that the adiabatic cyclic variation of $r_{\Omega}$ and $\kappa_{\Omega}$ encloses an
instantaneous Floquet EP; (b,e) The Floquet frequency is $\omega=52$ and it is chosen in a way that the EP is in the vicinity of the the adiabatic 
cyclic variation of $r_{\Omega}$ and $\kappa_{\Omega}$; (c,f) The Floquet frequency is $\omega=80$. In this case the EP is far away from the 
parameter domain where the  adiabatic cyclic variation of $r_{\Omega}$ and $\kappa_{\Omega}$ occurs. In all cases the fixed adiabatic parameter variations $r_{\Omega}$ and $\kappa_{\Omega}$ 
are $r_{\Omega}(t)=r_0 \sin(\Omega t)$ 
and  $\kappa_{\Omega}(t)=\kappa_0+ \kappa_1\cos(\Omega t)$  where $r_0=0.1$,  $\Omega=0.01$,
$\kappa_0=4$  and $\kappa_1=2$.
The other parameters are $\gamma=1$ and $F=100$. }
\label{fig2}
\end{center}
\end{figure}

In Fig.~\ref{fig2} we report the evolution of eigenvalues  $\lambda_{+}(t)$ (red lines) and $\lambda_{-}(t)$ (black lines) together with the evolution 
of the corresponding $p(t)$ (red circles and black diamonds) for three different values of the Floquet driving frequency $\omega=50, 52$ and $80$. The 
upper row corresponds to the real part of these quantities while their imaginary part is reported in the lower row. The parameters used in these
simulations are $\gamma=1$ and $F=100$, while the slow varying parameters $r_{\Omega}, \kappa_{\Omega}$ have been chosen to change as $r_{\Omega}(t)=r_0 \sin(\Omega t)$ 
and  $\kappa_{\Omega}(t)=\kappa_0+ \kappa_1\cos(\Omega t)$  where $r_0=0.1$,  $\Omega=0.01$,
$\kappa_0=4$  and $\kappa_1=2$. For all 
$\omega-$values presented in Fig.~\ref{fig2} the rate and the direction of evolution of the adiabatic circle remains the same. In Figs.~\ref{fig2}a,d
the EP is inside the circle while in Figs. \ref{fig2}b,e is in the proximity of it. In both cases, we find a state-flip as predicted from the analysis of
the theoretical model Eq. (\ref{eq: main}). In contrast, in Figs.~\ref{fig2}c,f the Floquet frequency $\omega$ is such that it has re-allocate the EP 
far away from the adiabatic circle. In this case we do not observe a state-flip. Instead the system remains at the same state as the original one 
at  the end of the cycle at $t=T_{\Omega}$.

{\it Conclusions--} We consider the state evolution of a non-Hermitian system which is exposed to two driving schemes with strong time-scale
separation. We have introduced the notion of Floquet state-flip due to adiabatic encircling of instantaneous Floquet EP singularities. Then we 
have used the extra degree of freedom that the Floquet driving is offering in order to re-organize the position of EP with respect to an adiabatic 
cycle associated with a slow variation of two additional parameters of the Hamiltonian. This EP re-organization leads to a tailoring of the topological 
features of the adiabatic cycle and allow us a state-flip reconfigurability. It will be interesting to realize this Floquet protocol using existing 
experimental platforms \cite{fred}.


\begin{thebibliography}{99}
\bibitem{note1}To derive Eq. (\ref{eq: main2}) from Eq. (\ref{eq: main}) we have assumed that terms ${\cal O}(\Omega)$ can be eliminated in the adiabatic 
limit of $\Omega\rightarrow 0$. This approximation is used only for the sake of simplicity and an alternate route without such an approximation can lead to 
a solution with qualitatively identical results.
\bibitem{note2} $C_{1}e^{\imath\nu\pi}-C_{2}\imath\pi=0$ is an exception,
however, it never exactly happen numerically. A tiny nonzero value
will be amplified extremely fast by the exponential growing factor
$e^{\nu\eta}$ in the adiabatic limit.
\bibitem{note3} Note that Hamiltonian Eq. (\ref{eq: main2}) can be considered a starting point for the analysis of a state-flip due to an adiabatic encircling 
of an EP associated with a non-Hermitian Hamiltonian without any Floquet driving. Therefore all our conclusions apply in the latter framework as well.
\end{thebibliography}
\end{document}